# Fair Isaac Technical Paper

**Subject:** **A Quadratic Programming Solution to the FICO Credit Scoring Problem**

**From:** **Bruce Hoadley (BCH – A34  X27051)**

**Date:** **July 14, 2000**

## Abstract

The classic Fair, Isaac score development problem is to find a score engineered scorecard that maximizes divergence. Over the years, Fair, Isaac has developed excellent solutions to this problem, but we have never solved it exactly – until now. INFORM9, INFORM11, INFORM12 (ITP), and INFORM13 (References [1], [5], [6], [7]) are all based on heuristic algorithms that find excellent solutions, but not exact solutions. We discovered this while developing the range engineering part of INFORM13. The heuristic Hooke-Jeeves algorithm was getting caught in high dimensional wedges and giving solutions that failed certain sanity checks. This stimulated Gerald Fahner and myself to look for alternative ways to solve the problem.

What we have discovered is that many of the classic Fair, Isaac score development problems can be solved by either a single quadratic program or a short (e.g., length 5) sequence of quadratic programs. In today's computing environment, this is attractive, because modern quadratic programming algorithms are very fast. For a large score engineered scorecard (171 attributes, 60 linear equality constraints, and 106 pattern constraints), the solving quadratic program ran in 7 seconds on my new Dell laptop using MATLAB.

The purpose of this paper is to describe how to solve five different score engineered score development problems with quadratic programming. The problems are called

1. Classic problem

2. Penalized classic problem

3. Non-zero in-weighting problem

4. Range engineering problem

5. Score engineered regression problem.

Problems 1, 2, and 5 can be solved exactly with a single quadratic program. Problems 3 and 4 require short sequences of quadratic programs.

For a fraud score test case, I provide all of the MATLAB code for solving the five problems, and provide the resulting score weights and divergences.

For the first problem, I compare my solution to two other solutions to the same problem. The first alternative solution is based on the Hooke-Jeeves version of INFORM-*NLP* (also known as INFORM13), and the second alternative solution is based on SAS quadratic programming, which was developed by Gerald Fahner.

The MATLAB and SAS versions of quadratic programming give virtually the same solution, and both yield higher divergence than INFORM13.

# Table of Contents

# 1. Introduction

## 1.1. Score formula

In this paper I consider a score of the form

$$Score = \sum_{j=1}^{p} S_j X_j,$$

Score = $\sum_{j=1}^{p} S_j X_j$

where the $X_j's$ are arbitrary numerical predictors and the $S_j's$ are score weights. Of course in multiple regression theory the $S_j's$ are called regression coefficients and in linear discriminant analysis the score is called a linear discriminant. In the case where the $X_j's$ are attribute indicator variables, the score is a scorecard. In INFORM-*NLP* (formerly called INFORM13) theory the score was transformed to so-called $(\mathbf{S}, \mathbf{X})$ space (see Reference [1]), where the score was also of the above form.

The score can be written in matrix notation as

$$Score = \mathbf{S'} * \mathbf{X},$$

where

$$\mathbf{S'} = (S_1, ..., S_p)$$
$$\mathbf{X'} = (X_1, ..., X_p).$$

I use bold to indicate a matrix or vector and use $*$ to indicate ordinary matrix multiplication.

## 1.2. Score moments

One problem considered in this paper is to maximize the divergence of $Score$ subject to a large set of score engineering constraints. In order to describe this problem mathematically, I need to introduce some notation.

The relevant score moments are

$$\mu_G = E[Score\,|\,G] = E[S' * X\,|\,G] = S' * E[X\,|\,G] = S' * MG$$

$$\mu_B = E[Score\,|\,B] = S' * MB$$

$$\sigma_G^2 = V[Score\,|\,G] = V[S' * X\,|\,G] = S' * Cov[X\,|\,G] * S = S' * CG * S$$

$$\sigma_B^2 = V[Score\,|\,G] = S' * CB * S$$

$$\sigma^2 = S' * \left[\frac{CG + CB}{2}\right] * S = S' * C * S\,.$$

The above expectations and variances are computed with respect to the observation weights (e.g., the sample weights) , which impose a probability distribution over the observations.

The divergence of the score is

$$Div[Score] = \frac{(\mu_G - \mu_B)^2}{\sigma^2}$$

$$= \frac{(S' * MG - S' * MB)^2}{S' * C * S}$$

$$= \frac{(S' * (MG - MB))^2}{S' * C * S}$$

$$= \frac{(d' * S)^2}{S' * C * S}\,.$$

This is a ratio of quadratic functions of $S$.

## 1.3. Score engineering

At Fair, Isaac, scores are developed under what is called score engineering. For scorecards, score engineering includes centering constraints, no-inform constraints, cross restrictions, grouping restrictions, in weighting, pattern constraints, score weight size constraints, and weight of evidence scale. All of these score engineering constraints amount to constraints on the score weights, $S$. Here I write these constraints in mathematical notation.

The centering, no-inform, cross restrictions, grouping restrictions, and in-weighting constraints can be defined by a set on linear equality constraints in $S$. In matrix notation these can be written as

$$Ac * S = bc.$$

The pattern constraints are a set of linear inequality constraints, which can be written as

$$Ap * S \leq bp.$$

The size constraints are a set of linear inequality constraints, which can be written as

$$l \leq S \leq u.$$

The two types of linear inequality constraints are specified separately, because they will be treated differently by the score development algorithm.

The weight of evidence scale constraint is the non-linear equality constraint

$$\sigma^2 = \mu_G - \mu_B$$

or

$$S' * C * S = d' * S.$$

This is a quadratic equality constraint in $S$.

## 1.4. Fair Isaac score development problem

The Fair, Isaac score development problem can now be expressed in mathematical programming language. It is to

Find $S$ to

Maximize $\dfrac{(d' * S)^2}{S' * C * S}$

Subject to :

$Ac * S = bc$

$S' * C * S = d' * S$

$Ap * S \leq bp$

$l \leq S \leq u.$

As it stands, this is not a quadratic program. The objective function is a ratio of quadratic functions – not quadratic. Most of the constraints are linear – except for the weight of evidence constraint, which is non-linear. So it is a non-linear programming problem.

I find it fascinating that this problem has never been optimally solved at Fair, Isaac – until recently. The current version of the production score development system is based on the INFORM11 weights engine. Using linear programming, INFORM11 produces a score, which is a sufficient statistic for log odds under proprietary assumptions. INFORM11 usually produces a score with good divergence, but it is not maximum divergence.

The Inform Test Platform (ITP) and the SAS version of INFORM-*NLP* use the Hooke-Jeeves pattern search algorithm to solve the classic Fair, Isaac problem, which is stated above. However, recent research has shown that the Hooke-Jeeves pattern search algorithm does not always precisely maximize divergence. This is because the algorithm is a heuristic and can get stuck in small wedges of score weight space. INFORM-*NLP* usually yields a higher divergence than INFORM11, but it is not always maximum divergence.

One of the aims of this paper is to show how to solve the Fair, Isaac score development problem exactly using quadratic programming.

## 2. Classic Problem

### 2.1. Mathematical formulation of the classic problem

In most applications of the Fair, Isaac problem, we have the conditions

$$bc = 0$$
$$bp = 0$$
$$l = -\infty$$
$$u = +\infty \, .$$

Clearly the centering constraints have zero on the right hand side. On the weight of evidence scale, the no-inform constraint has zero on the right hand side. A typical cross restriction or group restriction is of the form $S_i = S_j$, which can be written as $S_i - S_j = 0$.

The only exception to $bc = 0$ is non-zero in weighting, which is somewhat rare.

A typical pattern constraint is of the form $S_i \leq S_j$, which can be written as $S_i - S_j \leq 0$.

I call this typical case, the

**Classic problem**

Find $S$ to

Maximize $\dfrac{(d' * S)^2}{S' * C * S}$

Subject to :
$$Ac * S = 0$$
$$S' * C * S = d' * S$$
$$Ap * S \leq 0 .$$

One thing that makes this problem difficult is the non-linear weight of evidence constraint. The role of this constraint is to keep the score weights on a standard scale. Without this constraint, there would be no unique solution to this problem, because divergence is invariant to a multiplication of the score weights by a constant.

But there is an alternative way to constrain the score weights to be on some reasonable scale. Just use the simple linear constraint $d' * S = \delta$, where $\delta$ is an apriori guess at the maximum divergence associated with the optimal solution. Then the problem becomes

**Alternative to classic problem**

Find $S$ to

Maximize $\dfrac{(d' * S)^2}{S' * C * S}$

Subject to :
$$Ac * S = 0$$
$$d' * S = \delta$$
$$Ap * S \leq 0 .$$

This problem maximizes divergence on some other scale that is an approximation to the weight of evidence scale.

This problem can be re-expressed as

Find $S$ to

Maximize $\dfrac{\delta^2}{S' * C * S}$

Subject to :

$Ac * S = 0$

$d' * S = \delta$

$Ap * S \leq 0$ .

And this can be re-expressed as

## Classic quadratic program

Find $S$ to

Minimize $S' * C * S$

Subject to :

$Ac * S = 0$

$d' * S = \delta$

$Ap * S \leq 0$ ,

which is a quadratic program.

For a large sized scorecard, this quadratic program can be solved in about 7 seconds on my laptop using MATLAB.

This quadratic program maximizes divergence on a scale defined by $d' * S = \delta$, which is only an approximation to the weight of evidence scale. It turns out that if I transform the solution to the classic quadratic program to the weight of evidence scale, then I get a solution to the classic problem. I state this as a theorem.

## Theorem 1

Let $T$ be a solution to the classic quadratic program, and let $W = \beta T$ be its transformation to a weight of evidence scale. Then $W$ is a solution to the classic problem.

## *Proof*

Let $S$ be the solution to the classic problem. And let

$$U = \left(\dfrac{\delta}{d'S}\right)S = \alpha S.$$

In Lemma 1 below, I show that $W$ is feasible for the classic problem. Hence,

$$Div(W) \leq Div(S). \tag{1}$$

In Lemma 2 below, I show that $U$ is feasible for the classic quadratic program. Hence,

$$Div(U) \leq Div(T).$$

Since divergence is invariant to linear transformations, we have

$$Div(U) = Div(S)$$
$$Div(T) = Div(W).$$

Hence

$$Div(S) \leq Div(W). \tag{2}$$

Equations (1) and (2) imply

$$Div(W) = Div(S),$$

which proves Theorem 1.

### _Q.E.D_

**Lemma 1**

$W$ is feasible for the classic problem.

_Proof_

Define $W = \beta T$, where

$$\beta = \frac{d' * T}{T' * C * T}.$$

Note that

$$Ac * W = Ac * \beta T = \beta(Ac * T) = 0 \quad .$$

Next note that

$$W' * C * W = (\beta T)' * C * (\beta T) = \beta\beta(T' * C * T)$$
$$= \beta\left(\frac{d' * T}{T' * C * T}\right)(T' * C * T) = \beta(d' * T)$$
$$= d' * W.$$

And finally note that

$$Ap * W = Ap * \beta T = \beta(Ap * T) \leq 0 \quad .$$

So the score weight vector, $W = \beta T$, is on the weight of evidence scale, it satisfies the linear equality constraints, and it satisfies the linear inequality constraints.

_Q.E.D_

**Lemma 2**

$U$ is feasible for the classic quadratic program.

_Proof_

First note that

$$Ac * U = Ac * \alpha S = \alpha(Ac * S) = 0 \quad .$$

Next note that

$$d' * U = d' * \alpha S = d' * \left(\frac{\delta}{d'S}\right) S = \delta.$$

And finally note that

$$Ap * U = Ap * \alpha S = \alpha(Ap * S) \leq 0 \quad .$$

_Q.E.D_

In summary, I have shown that in order to solve the classic problem, you need only solve the classic quadratic program and transform the solution to the weight of evidence scale. This takes 7 seconds on my PC using MATLAB.

## 2.2. MATLAB quadratic programming

The general quadratic program solved by MATLAB is

Find $S$ to

Minimize $\left(\dfrac{1}{2}\right) S' * H * S + f'S$

Subject to :

$Aeq * S = beq$

$A * S \leq b$

$l \leq S \leq u$ .

## 2.3. MATLAB formulation of classic quadratic program

In the previous section I argued that the classic problem could be reduced to the classic quadratic program

Find $S$ to

Minimize $S' * C * S$

Subject to :

$Ac * S = 0$

$d' * S = \delta$

$Ap * S \leq 0$ ,

So the matrices in the general form of the MATLAB quadratic program are

$H = 2C$

$f = f0$ , where $f0 = 0$

$Aeq = \begin{bmatrix} d' \\ Ac \end{bmatrix}$

$beq = \begin{bmatrix} \delta \\ 0 \end{bmatrix}$

$A = Ap$

$b = bp,$ where $bp = 0$

$l = lb,$ where $lb = -\infty$

$u = ub,$ where $ub = +\infty$

## 2.4. Fraud case study

**Data**

For the case study, I use the high-risk fraud data used by Nina Shikaloff and Gerald Fahner in their INFORM-*NLP* studies. These data were also used in my previous fraud studies reported in [3] and [4].

The SAS data set that I started with is called iphr_i and is located in

$$/\text{fico}/\text{prod}/\text{ap}/\text{score}/\text{rde}/\text{znxs5}/\text{bldcard}/\text{unix}/\text{db}$$

The SAS program that I used to convert the relevant variables in the SAS data set to a UNIX text file is given in Appendix 1.

**INFORM-*NLP* scorecard**

Nina Shikaloff has applied the SAS version of INFORM-*NLP* to this data. This version of INFORM-*NLP* is based on the Hooke-Jeeves pattern search algorithm. Since this is a heuristic, it finds an excellent solution, but not necessarily the optimal solution. The scorecard is shown in Appendix 2. This large scorecard involves a heavy use of score engineering, so it is a good case study. The characteristics and attributes used are shown in this scorecard.

The attribute indicator variables in the SAS data set are of the form $vc\_a$, where $c = 1,2,...,25$ (varies over characteristics) and $a = 0,1,..., A_c - 1$, where $A_c$ is the number of attributes for characteristic $c$. For every characteristic, $a = 0$ corresponds to the NO INFORMATION attribute, which is always listed as the last attribute, even when it has no counts.

In Appendix 1, the characteristics are listed in the same order as they appear in the scorecard. And the attribute indicator variables are also ordered in the same order as they appear in the scorecard. So, for example, attribute indicator variables v1_1 - v1_6, v1_0 are associated with char170. The complete list is:

| Attribute Indicator Variables | Corresponding Characteristic |
|---|---|
| v1_1-v1_6    v1_0 | 170 |
| v2_1-v2_6    v2_0 | 191 |
| v3_1-v3_6    v3_0 | 193 |
| v4_1-v4_12   v4_0 | 211 |
| v5_1-v5_7    v5_0 | 260 |
| v6_1-v6_10   v6_0 | 320 |
| v7_1-v7_3    v7_0 | 330 |
| v8_1-v8_6    v8_0 | 380 |
| v9_1-v9_3    v9_0 | 471 |
| v10_1-v10_2  v10_0 | 503 |
| v11_1-v11_6  v11_0 | 533 |
| v12_1-v12_6  v12_0 | 635 |
| v13_1-v13_3  v13_0 | 665 |
| v14_1-v14_7  v14_0 | 710 |
| v15_1-v15_3  v15_0 | 830 |
| v16_1-v16_5  v16_0 | 835 |
| v17_1-v17_4  v17_0 | 840 |
| v18_1-v18_4  v18_0 | 843 |
| v19_1-v19_2  v19_0 | 860 |
| v20_1-v20_3  v20_0 | 870 |
| v21_1-v21_15 v21_0 | 950 |
| v22_1-v22_4  v22_0 | 960 |
| v23_1-v23_5  v23_0 | 961 |
| v24_1-v24_9  v24_0 | 962 |
| v25_1-v25_9  v25_0 | 965 |

### 2.5. MATLAB code for the classic quadratic program

As described above, in order to set up the classic quadratic program, I need to first compute the matrices

$$C, d, Ac, Ap, f0, bp, lb, ub.$$

### Computation of $C$

The SAS program in Appendix 1 created a UNIX text file called frdata, which had 14,000 rows and 206 variable – space delimited. This UNIX text file was converted to a PC text file called Frdata.txt. The MATLAB command for putting this into a $14{,}000 \times 206$ MATLAB matrix called Frdata is

```
load Frdata.txt;
```

In the matrix Frdata, the attribute indicator variables are variables 36 through 206. The MATLAB command for creating the overall design matrix is

```
X=Frdata(:,36:206);
```

<u>Warning</u>: Unfortunately, this `X` is related to, but not exactly the same as the random vector of prediction variables, $X$, which appears in the theory.

The indicator variable for non-fraud is variable 32. So the MATLAB command for creating the overall performance vector (a $14{,}000 \times 1$ vector) is

```
y=Frdata(:, 32);
```

The serial number is variable 3, so the MATLAB command for creating the serial number vector is

```
sn=Frdata(:, 3);
```

The validation sample corresponds to observation with serial numbers 1,4 and 8. So the MATLAB commands for creating the development and validation performance variable are

```
yv=y(((sn==1)|(sn==4)|(sn==8)));
yd=y(~((sn==1)|(sn==4)|(sn==8)));
```

Similarly, to create the design matrix for the development sample and the validation sample, use the MATLAB commands

```
Xd=X(~((sn==1)|(sn==4)|(sn==8)),:);
Xv=X(((sn==1)|(sn==4)|(sn==8)),:);
```

To create the development and validation design matrices for the goods and bads, the MATLAB commands are

```
XdG=Xd(yd==1,:);
XdB=Xd(yd==0,:);
XvG=Xv(yv==1,:);
XvB=Xv(yv==0,:);
```

To compute the development average covariance matrix, the MATLAB command is

```
C=(Cov(XdG)+Cov(XdB))/2;
```

## Computation of $d$

In MATLAB, this is easy. The command is

```
d=(mean(XdG)–mean(XdB))';
```

## Computation of $Ac$

Each row of $Ac$ corresponds to a score engineering equality constraint. In this problem there are 25 centering constraints, 25 no inform constraints, and 9 cross restriction constraints, for a total of 59. The number of score weights is 171. So $Ac$ is a $59 \times 171$ matrix.

The first step in the process is to create a $59 \times 171$ matrix of zeros. This is done with the MATLAB command

```
Ac=zeros(59,171);
```

### Centering constraints

Now I work on the first 25 rows of $Ac$, which are the centering constraints. There is one centering constraint for each characteristic. Define a vector, called high, of 25 score weight indices, which give the last attribute of each characteristic. For example, the last score weight index of characteristic 170 is 7 and the last score weight index of characteristic 191 is 14, etc. The MATLAB command is

```
high = [7 14 21 34 42 53 57 64 68 71 78 . .
.
        85 89 97 101 107 112 117 120 . . .
        124 140 145 151 161 171];
```

Next define a vector, called low, of 25 score weigh indices, which give the first attribute of each characteristic. To do this, the MATLAB commands are

```
low=ones(1,25);

for i=2:25

    low(i)=high(i-1)+1;

end
```

The centering constraints involve attribute weights. Initially, all of these weights can be put into one vector with the MATLAB command

```
e=mean(XdG)+mean(XdB);
```

Now we can replace the first 25 rows of $Ac$ with the correct values, which are given by the MATLAB commands

```
for i=1:25

    for k=low(i):high(i)

        Ac(i,k)=e(k) ;

    end

end
```

<u>No Information constraints</u>

The next 25 rows of $Ac$ contain the left-hand side of the no information constraints. A no information constraint just sets a no information score weight equal to zero. The MATLAB commands for the next 25 rows are

```
for i=1:25

    Ac(i+25,high(i))=1;

end
```

<u>Cross restrictions</u>

Examination of the cross restrictions in the scorecard in Appendix 2 leads to the following MATLAB code for setting up rows 51 through 59 of $Ac$.

```
Ac(51,1)=1;
Ac(52,8)=1;
Ac(53,58)=1;
Ac(54,65)=1;
Ac(55,90)=1;
Ac(56,91)=1;
Ac(57,69)=1;
Ac(57,98)=-1;
Ac(58,69)=1;
Ac(58,118)=-1;
Ac(59,69)=1;
Ac(59,121)=-1;
```

## **Computation of** $Ap$

From the scorecard in Appendix 2 we can see that there are 106 pattern constraints. Each pattern constraint corresponds with one row of the matrix $Ap$, so $Ap$ is a $106 \times 171$ matrix. We initialize $Ap$ with all zeros via the MATLAB command

```
Ap=zeros(106,171);
```

Each pattern constraint can be written in the form $(+1)S_j + (-1)S_k \le 0$ or $(-1)S_j + (+1)S_k \le 0$, where $j < k$. There are 106 values or $j$, which I put into the MATLAB vector

```
j=[2 3 4 5 9 10 11 12 16 17 18 19 24 . . .
   25 26 27 28 29 30 31 32 36 37 38 . . .
   39 40 43 44 45 46 47 48 49 50 51 . . .
   54 55 59 60 61 62 66 69 69 73 74 . . .
   75 76 80 81 82 83 86 87 92 93 94 . . .
   95 98 99 102 103 104 105 108 109 . . .
   110 118 121 122 125 126 127 128 . . .
   129 130 131 132 133 134 135 136 . . .
   137 138 141 142 143 147 148 149 . . .
   152 153 154 155 156 157 158 159 . . .
   162 163 164 165 166 167 168 169];
```

Associated with each $j$ is a value of $k$. In most cases, $k = j + 1$, but there are exceptions. The MATLAB code for computing the vector of $k's$ is

```
k=j+1;
k(13)=31;
k(14)=30;
k(44)=71;
```

Associated with each $j$ is a coefficient, which is either (+1) or (-1) (see above discussion). The coefficients associated with the $j's$ are given in the MATLAB vector, `a`, which is computed by the following MATLAB code.

```
a=ones(1,106);
a(1:4)=-1;
a(22:41)=-1;
a(43:84)=-1;
```

The coefficients associated with the $k's$ are given in the MATLAB vector, `na`, which is just the negative of `a`, and is computed by the following MATLAB code.

```
na=-a;
```

These vectors can now be used to compute the $Ap$ matrix as follows

```
for i = 1:106
        Ap(i,j(i))=a(i);
        Ap(i,k(i))=na(i);
end
```

## Computation of $f0, bp, lb, ub$

```
f0=zeros(171,1);
bp=zeros(106,1);
lb=(-inf)*ones(171,1);
ub=inf*ones(171,1);
```

## Matrices for the classic quadratic program

As we saw above, the matrices needed for the quadratic classic problem are

$$H = 2C$$
$$f = 0$$
$$Aeq = \begin{bmatrix} d' \\ Ac \end{bmatrix}$$
$$beq = \begin{bmatrix} \delta \\ 0 \end{bmatrix}$$
$$A = Ap$$
$$b = 0$$
$$l = -\infty$$
$$u = +\infty .$$

In this paper, I will be describing many application of the MATLAB quadratic programming algorithm. The MATLAB notation for these matrices for this first problem is `H1, f1, Aeq1, beq1, A1, b1, l1,` and `u1`.

The MATLAB code for computing these matrices is

```
H1=2*C;
f1=f0;
Aeq1=[d';Ac];
beq1=[1.753;zeros(59,1)];
A1=Ap;
b1=bp;
l1=lb;
u1=ub;
```

Note that I chose $\delta = 1.753$.

### Initial solution for the quadratic program

To run the MATLAB quadratic program, you need an initial solution. I suspect that it is a good idea to have a pretty good initial solution. Here is how I generated a pretty good initial solution.

I first ran the quadratic program with no pattern constraints with the MATLAB code

```
x01=quadprog(H1,f0,[],[],Aeq1, beq1, lb,
ub);
```

The variable `x01` is a score weight vector. I then heuristically adjusted this score weight vector so that it was feasible for the pattern constraints.

## Solving the quadratic program

The MATLAB code for solving the quadratic program is

```
S1=quadprog(H1,f0,Ap,bp,Aeq1,beq1,lb,ub,x01)
;
```

## Transformation to weight of evidence scale

As we have discussed, this solution is not quite on a weight of evidence scale. You convert it to the weight of evidence scale via the MATLAB code

```
scr1dG=XdG*S1;

scr1dB=XdB*S1;

beta1=(2*(mean(scr1dG)- mean(scr1dB)))/ . .
.
          (cov(scr1dG)+cov(scr1dB));

S1=beta1*S1;
```

### 2.6. Fraud solution to the classic quadratic program

The solution is given in Appendix 2. The development divergence is 1.753. The development divergence for the Hooke-Jeeves version of INFORM-*NLP* is 1.732, which is less than 1.753. This proves that the Hooke-Jeeves version of INFORM-*NLP* does not maximize divergence even though it is designed to do so. The reason is the heuristic nature of the Hooke-Jeeves pattern search algorithm.

The quadratic program also produced a score that has more validation divergence than the Hooke-Jeeves INFORM-*NLP* score. This would not always be the case, but it was true for this example.

Gerald Fahner has developed SAS software for maximizing divergence using SAS quadratic programming. His solution is given as the second score in Appendix 2. As you can see, Gerald's solution is virtually the same as mine.

## 3. Penalized Classic Problem

### 3.1. Mathematical formulation of the penalized classic problem

The penalized version of the classic problem is

## Penalized classic quadratic program

Find $S$ to

Minimize $\quad S' * C * S + \dfrac{\lambda}{p} S' * S$

Subject to :

$\quad Ac * S = 0$

$\quad d' * S = \delta$

$\quad Ap * S \leq 0$ .

This can be re-expressed as

Find $S$ to

Minimize $\quad S' * \left( C + \dfrac{\lambda}{p} I \right) * S$

Subject to :

$\quad Ac * S = 0$

$\quad d' * S = \delta$

$\quad Ap * S \leq 0$ ,

where $I$ is a $p \times p$ identity matrix. As you can see, this is also a quadratic program.

### 3.2. MATLAB formulation of the penalized classic quadratic program

For the penalized classic quadratic program, the matrices in the general form of the MATLAB quadratic program are

$$H = 2\left(C + \frac{\lambda}{p} I\right)$$

$$f = 0$$

$$Aeq = \begin{bmatrix} d' \\ Ac \end{bmatrix}$$

$$beq = \begin{bmatrix} \delta \\ 0 \end{bmatrix}$$

$$A = Ap$$

$$b = 0$$

$$l = -\infty$$

$$u = +\infty .$$

### 3.3. MATLAB code for the penalized classic quadratic program

The MATLAB notation for these matrices for the penalized classic quadratic program is `H2, f2, Aeq2, beq2, A2, b2, l2,` and `u2`.

The MATLAB code for computing these matrices is

```
H2=2*(C+(.095/171)*eye(171));
f2=f0;
Aeq2=Aeq1;
beq2=beq1;
A2=Ap;
b2=bp;
l2=lb;
u2=ub);
```

The only matrix that changes for the penalized classic quadratic program is the $H$ matrix. Note that I chose $\lambda = .095$. This is the value that maximized divergence on the validation sample. It was found by a line search in $\lambda$ space.

The MATLAB code for solving the penalized classic quadratic program is

```
S2=quadprog(H2,f0,Ap,bp,Aeq1,beq1,lb,ub,x01)
;
```

## 3.4. Fraud solution to the penalized classic quadratic program

As we have discussed, this solution is not quite on a weight of evidence scale, but can be transformed to the weight of evidence scale in the usual way.

The solution is given in Appendix 2. The development divergence is 1.752, which is smaller than the development divergence for Score 1 – the un-penalized case. However, the validation divergence is slightly larger than for score 1, which is what was expected with an optimal choice of $\lambda$.

For many of the attributes, the score weights for Score 2 are closer to zero than the score weights for Score 1. This is what I expected. However, the number of attributes, for which this was not true, surprised me.

# 4.0 Non-zero In-weighting Problem

## 4.1. Mathematical formulation of the non-zero in-weighting problem

A non-zero in-weight is of the form $S_j = IW_j$, where $IW_j \neq 0$. In this case, the classic Fair, Isaac problem can not be solved by a single quadratic program. In this section I propose an iterative quadratic programming solution to the problem.

The non-zero in-weighting problem can be formulated as

**Non-zero in-weighting problem**

Find $S$ to

Maximize $\dfrac{(d' * S)^2}{S' * C * S} - \dfrac{\lambda}{p} S' * S$

Subject to :

$Ai * S = IW$

$Ac * S = 0$

$S' * C * S = d' * S$

$Ap * S \leq 0$ .

This problem can be re-expressed by plugging the weight of evidence constraint into the objective function. The result is

Find $S$ to

Maximize $d' * S - \dfrac{\lambda}{p} S' * S$

Subject to :
$$Ai * S = IW$$
$$Ac * S = 0$$
$$S' * C * S = d' * S$$
$$Ap * S \leq 0 \,.$$

This can be converted to a minimization problem by changing the sign of the objective function. The result is

Find $S$ to

Minimize $- d' * S + \dfrac{\lambda}{p} S' * S$

Subject to :
$$Ai * S = IW$$
$$Ac * S = 0$$
$$S' * C * S = d' * S$$
$$Ap * S \leq 0 \,.$$

This is almost a quadratic program. The fly in the ointment is the quadratic equality constraint $S' * C * S = d' * S$.

It is well known (private communication from Bob Oliver) that under regularity conditions, this problem can be solved by putting the quadratic constraint into the objective function with a Lagrange multiplier, and then solving an iterative sequence of quadratic programs.

For a fixed value of the Lagrange multiplier, $\phi$, you get the quadratic program

**<u>Non-zero in-weighting quadratic program</u>**

Find $S$ to

Minimize $-d' * S + \dfrac{\lambda}{p} S' * S + \phi(S' * C * S - d' * S)$

Subject to :
$$Ai * S = IW$$
$$Ac * S = 0$$
$$Ap * S \leq 0 \,.$$

Let $S(\phi)$ be the solution to this quadratic program with a fixed $\phi$. The non-zero in-weighting problem can be solved by solving the non-linear equation

$$S(\phi)' * C * S(\phi) - d' * S(\phi) = 0.$$

This equation can be solved by standard line search techniques.

## 4.2. MATLAB formulation of the non-zero in-weighting quadratic program

The objective function for the non-zero in-weighting quadratic program can be re-written as

$$\phi S' * C * S + \frac{\lambda}{p} S' * S - (1 + \phi) d' * S$$

$$= S' * \left( \phi C + \frac{\lambda}{p} I \right) * S - (1 + \phi) d' * S$$

$$= \left( \frac{1}{2} \right) S' * 2 \left( \phi C + \frac{\lambda}{p} I \right) * S - (1 + \phi) d' * S \,.$$

So the matrices in the general form of the MATLAB quadratic program are

$$H = 2\left(\phi C + \frac{\lambda}{p} I\right)$$
$$f = -(1 + \phi)d$$
$$Aeq = \begin{bmatrix} Ai \\ Ac \end{bmatrix}$$
$$beq = \begin{bmatrix} IW \\ 0 \end{bmatrix}$$
$$A = Ap$$
$$b = 0$$
$$l = -\infty$$
$$u = +\infty.$$

### 4.3. MATLAB code for the non-zero in-weighting quadratic program

In this application $p = 171$ and I will take $\lambda = .095$, which was the optimal value for the penalized classic quadratic program.

The MATLAB notation for the matrices in the non-zero in-weighting quadratic program is `H3, f3, Aeq3, beq3, A3, b3, l3,` and `u3`.

The MATLAB code for computing the matrices, which are the same as for the classic quadratic program, is

```
A3=Ap;
b3=bp;
l3=lb;
u3=ub;
```

As an initial value of $\phi$ I take $\phi = .1$. So the MATLAB code for computing `H3` and `f3` is

```
H3=2*((.1)*C+(.095/171)*eye(171));
f3=-(1+.1)*d;
```

To compute `Aeq3` and `beq3` I first need to specify the non-zero in weighting. The non-zero in-weighting constraints are

$$S_2 = .5$$
$$S_{13} = .3.$$

For Characteristic 170, this increases the first positive score weight, and for Characteristic 191, this increases the last positive score weight.

The MATLAB code for computing `Ai is`

```
Ai=zeros(2,171);
Ai(1,2)=1;
Ai(2,13)=1;
```

The MATLAB code for computing `Aeq3` and `beq3 is`

```
Aeq3=[Ai;Ac];
beq3=[.5;.3;zeros(59,1)];
```

The MATLAB code for solving the non-zero in-weighting quadratic program is

```
S3=quadprog(H3,f3,Ap,bp,Aeq3,beq3,lb,ub,S2);
```

### 4.4. Fraud solution to the non-zero in-weighting quadratic program

The iterations of this quadratic program, over $\phi$ during the line search, went as follows:

| $\phi$ | $S3' * C * S3 - d' * S3$ |
|---|---|
| .1 | 36.45 |
| 2 | -.295 |
| 1.75 | -.262 |
| .75 | .341 |
| 1 | .019 |
| 1.025 | -.0012 |

The final score weights, $S3$, along with the development and validation divergences are shown in Appendix 2. As expected, the development divergence is slightly smaller due to the in weighting. Also note that $S_2$ and $S_{13}$ are in-weighted to the correct values of .5 and .3 respectively. And I verified that Score 3 is on the weight of evidence scale.

## 5. Range Engineering Problem

### 5.2 Mathematical formulation of the range engineering problem

A general formulation of the range engineering problem is

Find $S$ to

$$\text{Minimize } \sum_{j=1}^{p} R_j \left(S_j - T_j\right)^2 + \frac{\lambda}{p} S' * S$$

Subject to :

$$Ai * S = IW$$

$$Ac * S = 0$$

$$S' * C * S = d' * S$$

$$Ap * S \le 0$$

$$\frac{(d' * S)^2}{S' * C * S} \ge Div .$$

The parameter, $Div$, is a relaxation of the divergence obtained in Stage 1 of INFORM-NLP. If we take Score 3 to be the result of Stage 1, then the relaxation of the divergence can be something like

$$Div = (.95)(1.731)$$
$$= 1.64 .$$

The $T_j's$ are target score weights for the range engineering. The idea is to move the score weights towards these targets without losing too much divergence. If you want to shrink certain score weights as much as possible, then you could set their $T_j's$ to zero. If you want to expand certain positive score weights as much as possible (within reason), then you can set their $T_j's$ too be something like +3. If you want to expand certain negative score weights as much as possible (within reason), then you can set their $T_j's$ too be something like -3.

The $R_j's$ allow you to put more or less emphasis on engineering the individual score weights.

In this paper I will not address the issue of a user interface to these parameters, or even what recommended values they should take. It is clear that this mathematical approach gives the user a lot of flexibility in engineering the score weight ranges.

To express this problem in matrix notation, note that

$$\sum_{j=1}^{p} R_j \left(S_j - T_j\right)^2 = \left(S - T\right)' * R * \left(S - T\right),$$

where $R$ is a $p \times p$ diagonal matrix with $R_j$ as the $jth$ diagonal element. Further matrix algebra yields

$$\left(S - T\right)' * R * \left(S - T\right) = S'RS - 2T'RS + T'RT.$$

So the range engineering problem can be re-expressed as

> Find $S$ to
>
> Minimize $S'RS - 2T'RS + T'RT + \dfrac{\lambda}{p} S' * S$
>
> Subject to :
>
> $\quad Ai * S = IW$
>
> $\quad Ac * S = 0$
>
> $\quad S' * C * S = d' * S$
>
> $\quad Ap * S \leq 0$
>
> $\quad \dfrac{\left(d' * S\right)^2}{S' * C * S} \geq Div.$

The term $T'RT$ can be dropped from the objective function, because it is a constant. And the weight of evidence constraint can be used to simplify the divergence constraint. The result is

> Find $S$ to
>
> Minimize $S'RS - 2T'RS + \dfrac{\lambda}{p} S' * S$
>
> Subject to :
>
> $\quad Ai * S = IW$
>
> $\quad Ac * S = 0$
>
> $\quad S' * C * S = d' * S$
>
> $\quad Ap * S \leq 0$
>
> $\quad -d' * S \leq -Div.$

The objective function can now be re-arranged to yield

Find $S$ to

Minimize $S'\left(R + \dfrac{\lambda}{p}I\right)S - 2T'RS$

Subject to :

$Ai * S = IW$

$Ac * S = 0$

$S' * C * S = d' * S$

$Ap * S \leq 0$

$-d' * S \leq -Div$.

This is almost a quadratic program. The fly in the ointment is the quadratic equality constraint $S' * C * S = d' * S$.

It is well known (private communication from Bob Oliver) that under regularity conditions, this problem can be solved by putting the quadratic constraint into the objective function with a Lagrange multiplier, and then solving an iterative sequence of quadratic programs.

For a fixed value of the Lagrange multiplier, $\phi$, you get the quadratic program

Find $S$ to

Minimize $S'\left(R + \dfrac{\lambda}{p}I\right)S - 2T'RS + \phi(S' * C * S - d' * S)$

Subject to :

$Ai * S = IW$

$Ac * S = 0$

$Ap * S \leq 0$

$-d' * S \leq -Div$.

The objective function can be re-arranged to yield

**Range engineered quadratic program**

Find $S$ to

Minimize $S'\left(\phi C + R + \dfrac{\lambda}{p} I\right) S - (\phi d' + 2T'R)S$

Subject to :

$$Ai * S = IW$$
$$Ac * S = 0$$
$$-d' * S \leq -Div$$
$$Ap * S \leq 0 \ .$$

Let $S(\phi)$ be the solution to this quadratic program with a fixed $\phi$. The range engineering problem can be solved by solving the non-linear equation

$$S(\phi)' * C * S(\phi) - d' * S(\phi) = 0.$$

This equation can be solved by standard line search techniques.

## 5.2. MATLAB formulation of the range engineered quadratic program

The matrices in the general form of the MATLAB quadratic program are

$$H = 2\left(\phi C + R + \dfrac{\lambda}{p} I\right)$$
$$f = -(\phi d + 2RT)$$
$$Aeq = \begin{bmatrix} Ai \\ Ac \end{bmatrix}$$
$$beq = \begin{bmatrix} IW \\ 0 \end{bmatrix}$$
$$A = \begin{bmatrix} -d' \\ Ap \end{bmatrix}$$
$$b = \begin{bmatrix} -Div \\ 0 \end{bmatrix}$$
$$l = -\infty$$
$$u = +\infty \ .$$

## 5.3. MATLAB code for the range engineered quadratic program

In this application $p = 171$ and I will take $\lambda = .095$, which was the optimal value for the penalized classic quadratic program.

The MATLAB notation for the matrices in the range engineering quadratic program is `H4, f4, Aeq4, beq4, A4, b4, l4`, and `u4`.

The MATLAB code for computing the matrices, which are the same as for the classic quadratic program (problem 1), is

```
l4=lb;
u4=ub;
```

The MATLAB code for computing the matrices, which are the same as for the non-zero in-weighting quadratic program (problem 3), is

```
Aeq4=Aeq3;
beq4=beq3;
```

The MATLAB code for computing the matrices `A4` and `b4` is

```
A4=[–d';Ap];
b4=[–1.64;zeros(106,1)];
```

To compute `H4` and `f4` I need to first compute $R$ and $T$. For this application, I will only do range engineering on the first two characteristics; i.e., the first 14 attributes. I will make all 14 attributes equally important, so

$$R_j = \begin{cases} 1 & \text{for } j = 1,2,...,14 \\ 0 & \text{otherwise} \end{cases}.$$

I want to do range restriction for characteristic 170 and range expansion for characteristic 191. For characteristic 170, the target attribute score weights will be roughly one half of the optimal score weights; i.e., 0, .15, .08, -.03, -.13, -.44, and 0. For characteristic 191, the target attribute score weights will be roughly 1.5 times the optimal score weights; i.e., 0, -1.71, -1.55, -.86, -.003, .14, and 0.

The MATLAB code for computing `R` is

```
R=zeros(171,171);
for i=1:14
   R(i,i)=1;
end
```

The MATLAB code for computing `T` is

```
T=zeros(171,1);
T(1:14)=[0 .15 .08 −.03 −.13 −.44 0 . . .
         0 −1.71  −1.55 −.86 −.003 .14 0];
```

As an initial value of $\phi$ I take (based on some experience with this problem) $\phi = .019$. So the MATLAB code for computing `H4` and `f4` is

```
H4=2*((.019)*C+R+(.095/171)*eye(171));
f4=−((.019)*d+2*R*T);
```

The MATLAB code for solving the range engineered quadratic program is

```
S4=quadprog(H4,f4,A4,b4,Aeq3,beq3,lb,ub,S3);
```

## 5.4. Fraud solution to the range engineered quadratic program

The iterations of this quadratic program, over $\phi$ during the line search, went as follows:

| $\phi$ | $S4' * C * S4 - d' * S4$ |
|--------|--------------------------|
| .019   | .0120                    |
| .025   | -.00063                  |

The final score weights, $S4$, along with the development and validation divergences are shown in Table 1. As expected, the development divergence is 1.641, which is slightly above the lower bound of 1.64. Note that $S_2$ and $S_{13}$ are in-weighted to the correct values of .5 and .3 respectively. The score weight range for characteristic 170 has been restricted and the score weight range for characteristic 191 has been expanded. And I verified that score 4 is on the weight of evidence scale.

## 6. Score Engineered Regression Problem

### 6.1. Mathematical formulation of the score engineered regression problem

In a regression problem there is a dependent variable, $y$, which usually takes on a variety of numerical values. Revenue is an example of a dependent variable. For linear regression, the model fit to the dependent variable is of the form

$$E[y \mid X] = S_0 + \sum_{j=1}^{p} S_j X_j.$$

The score engineered regression problem can be formulated as

Find $S_0$ and $\boldsymbol{S}$ to

Minimize $\displaystyle \sum_{i=1}^{n} \left( y_i - \left( S_0 + \sum_{j=1}^{p} S_j x_{ij} \right) \right)^2 + \frac{\lambda}{p} \boldsymbol{S}' * \boldsymbol{S}$

Subject to :

$\quad \boldsymbol{Ai} * \boldsymbol{S} = \boldsymbol{IW}$

$\quad \boldsymbol{Ac} * \boldsymbol{S} = 0$

$\quad \boldsymbol{Ap} * \boldsymbol{S} \leq 0.$

Note that in the regression case, there is no weight of evidence scale. This is because the dependent variable, $y$, dictates the scale of the final model. To specify reasonable in weighting, you have to know about the natural scale of the score weights.

In the case where the $X_j's$ are attribute indicator variables, the scorecard part of the final model is $\displaystyle \sum_{j=1}^{p} S_j X_j.$

The score engineered regression problem can be put into matrix notation by defining some new matrices. Let

$$\beta' = \begin{bmatrix} S_0 & S' \end{bmatrix}$$

$$y' = \begin{bmatrix} y_1, y_2, \ldots, y_n \end{bmatrix}$$

$$Xr = \begin{bmatrix} 1 & x_{11} & x_{12} & \bullet & \bullet & x_{1p} \\ 1 & x_{21} & & & & x_{2p} \\ \bullet & \bullet & & x_{ij} & & \bullet \\ \bullet & \bullet & & & & \bullet \\ 1 & x_{n1} & x_{n2} & \bullet & \bullet & x_{np} \end{bmatrix}$$

$$Ir = \begin{bmatrix} 0 & 0 & 0 & 0 \\ 0 & 1 & 0 & 0 \\ 0 & 0 & 1 & 0 \\ 0 & 0 & 0 & 1 \end{bmatrix} \quad ((p+1) \times (p+1) \text{ matrix})$$

$$Air = \begin{bmatrix} 0 & Ai \end{bmatrix}$$

$$Acr = \begin{bmatrix} 0 & Ac \end{bmatrix}$$

$$Apr = \begin{bmatrix} 0 & Ap \end{bmatrix}.$$

In this notation, the score engineered regression problem is

Find $\beta$ to

Minimize $\left( y - Xr * \beta \right)' \left( y - Xr * \beta \right) + \dfrac{\lambda}{p} \beta' * Ir * \beta$

Subject to :

$Air * \beta = IW$

$Acr * \beta = 0$

$Apr * \beta \leq 0$.

The first part of the objective function can be expanded as follows:

$$\left( y - Xr * \beta \right)' \left( y - Xr * \beta \right)$$
$$= \beta' * Xr' * Xr * \beta - 2 * y' * Xr * \beta + y' * y.$$

Since $y' * y$ is a constant, it can be dropped from the objective function. The result is

**<u>Score engineered regression quadratic program</u>**

Find $\beta$ to

Minimize $\beta' * \left( Xr' * Xr + \dfrac{\lambda}{p} Ir \right) * \beta - 2 * y' * Xr * \beta$

Subject to :
$$Air * \beta = IW$$
$$Acr * \beta = 0$$
$$Apr * \beta \leq 0 \,.$$

## 6.2. MATLAB formulation of the score engineered regression quadratic program

For the score engineered regression quadratic program, the matrices in the general form of the MATLAB quadratic program are

$$H = 2 \left( Xr' * Xr + \frac{\lambda}{p} Ir \right)$$
$$f = -2 \left( Xr' * y \right)$$
$$Aeq = \begin{bmatrix} Air \\ Acr \end{bmatrix}$$
$$beq = \begin{bmatrix} IW \\ 0 \end{bmatrix}$$
$$A = Apr$$
$$b = 0$$
$$l = -\infty$$
$$u = +\infty \,.$$

## 6.3. MATLAB code for the score engineered regression quadratic program

The first step to computing these matrices is to compute
$y, Xr, Ir, Air, Acr, Apr,$ and $IW$.

I take the variable $y$ to be the indicator variable of non-fraud. The MATLAB vector for this variable is yd (see Section 2.5), where d stands for the development sample.

The MATLAB code for computing the rest of the preliminary matrices is

```
Xr=[ones(9907,1) Xd];
Ir=eye(172);
Ir(1,1)=0;
Air=[zeros(2,1) Ai];
Acr=[zeros(59,1) Ac];
Apr=[zeros(106,1) Ap];
IW=[.5;.3];
```

For the computation of `Xd,Ai,Ac,` and `Ap` see Sections 2.5, 4.3, 2.5, and 2.5 respectively.

The MATLAB notation for the matrices in the score engineered regression quadratic program is `H5, f5, Aeq5, beq5, A5, b5, l5,` and `u5.`

The MATLAB code for these matrices is

```
H5=2*(Xr'*Xr+(.095/171)*Ir);
f5=-2*(Xr'*yd);
Aeq5=[Air;Acr];
beq5=[IW;zeros(59,1)]
A5=Apr;
b5=bp;
l5=lb;
u5=ub;
```

To run the score engineered regression quadratic program, I need a starting solution. For this I will use

```
x05=[.5;S3];
```

The MATLAB code for solving the score engineered regression quadratic program is

```
beta5=quadprog(H5,f5,Apr,bp,Aeq5,beq5,lb,ub,x05);
```

### 6.4. Fraud solution to the score engineered regression quadratic program

The intercept term is `beta5(1)` = .499. The scorecard part of the regression formula is called $S5$ and is shown in Appendix 2. Note that the score weights tend to be smaller than for the other scores. This is because $S5$ is not on a weight of evidence scale. In fact you have to multiply $S5$ by 2.874 to transform it to a weight of evidence scale. However, $S5$ satisfies all of the score engineering constraints – including the non-zero in weighting.

The development divergence of $S5$ is only .926. This is partly due to the inappropriate in weighting, but it also shows that simple least squares regression is not the best way to solve the binary outcome problem.

# References

[1] Hoadley, Bruce, "A General Theory of Score Engineering for INFORM13," Fair, Isaac restricted Technical Paper, August 26, 1999.

[2] Hoadley, Bruce, "The Theory of INFORM 11," Fair, Isaac restricted Technical Notes, January 3, 1996.

[3] Hoadley, Bruce, "Boosting the Performance of Transaction Fraud Scores," Fair, Isaac Technical Paper, August 4, 1998.

[4] Hoadley, Bruce, "*RocBoost* - A New technology for Boosting the ROC Curve at an Operating Point," Fair, Isaac Technical Paper, December 16, 1998.

[5] Hoadley, Bruce, "INFORM13 – a Proposal for the Assimilation of Multiple Technologies into the INFORM12 Score Development Paradigm," Fair, Isaac Technical Monograph, January 28,1999.

[6] Hoadley, Bruce, "INFORM Moment Update Formulas for Arbitrary Numerical Prediction Variables," Fair, Isaac restricted Technical Paper, February 1, 1999.

[7] Hoadley, Bruce, "INFORM13 Weights Algorithm," Fair, Isaac restricted Technical Paper, April 1, 1999.

## Appendix 1. Program for Converting a SAS Data Set to a Unix Text File

```
libname bch v603
 "/fico/prod/ap/score/rde/znxs5/bldcard/unix/db" ;
data _null_ ;
     set bch.iphr_i ;
     recordno=_n_ ;
     blank=' ' ;
     file frdata recfm=d lrecl=805;
```

```
     put (recordno) (6.) (blank) ($2.) (acctnb) ($16.)
( serialn factin prfnum
char170 char191 char193 char211 char260 char320 char330
char380 char471 char503 char533 char635 char665 char710 char830 char835
char840 char843 char860 char870 char950 char960 char961 char962
char965) (28*15.5)
( v0_0 v0_1-v0_4
 v1_1-v1_6    v1_0
 v2_1-v2_6    v2_0
 v3_1-v3_6    v3_0
 v4_1-v4_12   v4_0
 v5_1-v5_7    v5_0
 v6_1-v6_10   v6_0
 v7_1-v7_3    v7_0
 v8_1-v8_6    v8_0
 v9_1-v9_3    v9_0
 v10_1-v10_2  v10_0
 v11_1-v11_6  v11_0
 v12_1-v12_6  v12_0
 v13_1-v13_3  v13_0
 v14_1-v14_7  v14_0
 v15_1-v15_3  v15_0
 v16_1-v16_5  v16_0
 v17_1-v17_4  v17_0
 v18_1-v18_4  v18_0
 v19_1-v19_2  v19_0
 v20_1-v20_3  v20_0
 v21_1-v21_15 v21_0
 v22_1-v22_4  v22_0
 v23_1-v23_5  v23_0
 v24_1-v24_9  v24_0
 v25_1-v25_9  v25_0) (176*2.) ;
run ;
```

# Appendix 2. Score Engineered Scorecards

| Char | Attribute | Att. # | Constraint | INFORM-NLP | GEF Max Div | S1 | S2 | S3 | S4 | S5 |
|------|-----------|--------|------------|------------|-------------|----|----|----|----|----|
| char170 | -9999999 | 1 | " = 0 " | 0.000 | 0.000 | 0.000 | 0.000 | 0.000 | 0.000 | 0 |
| char170 | 0-<5 | 2 | > 3 | 0.303 | 0.306 | 0.306 | 0.307 | 0.500 | 0.500 | 0.5 |
| char170 | 5-<25 | 3 | > 4 | 0.155 | 0.157 | 0.157 | 0.156 | 0.131 | 0.018 | -0.022 |
| char170 | 25-<35 | 4 | > 5 | -0.060 | -0.067 | -0.067 | -0.069 | -0.078 | -0.050 | -0.047 |
| char170 | 35-<300 | 5 | > 6 | -0.260 | -0.259 | -0.259 | -0.260 | -0.288 | -0.164 | -0.116 |
| char170 | 300-High | 6 | | -0.890 | -0.888 | -0.888 | -0.874 | -0.923 | -0.445 | -0.272 |
| char170 | NO INFORMATION | 7 | " = 0 " | 0.000 | 0.000 | 0.000 | 0.000 | 0.000 | 0.000 | 0 |
| char191 | -9999999 | 8 | " = 0 " | 0.000 | 0.000 | 0.000 | 0.000 | 0.000 | 0.000 | 0 |
| char191 | 0-<2 | 9 | < 10 | -1.133 | -1.152 | -1.150 | -1.144 | -1.294 | -1.712 | -0.435 |
| char191 | 2-<5 | 10 | < 11 | -1.086 | -1.084 | -1.088 | -1.038 | -1.144 | -1.551 | -0.372 |
| char191 | 5-<7 | 11 | < 12 | -0.625 | -0.626 | -0.630 | -0.573 | -0.693 | -0.861 | -0.29 |
| char191 | 7-<650 | 12 | < 13 | -0.005 | 0.000 | 0.000 | -0.002 | -0.134 | -0.111 | -0.181 |
| char191 | 650-High | 13 | | 0.101 | 0.096 | 0.096 | 0.095 | 0.300 | 0.300 | 0.3 |
| char191 | NO INFORMATION | 14 | " = 0 " | 0.000 | 0.000 | 0.000 | 0.000 | 0.000 | 0.000 | 0 |
| char193 | -9999999 | 15 | | 0.397 | 0.396 | 0.396 | 0.393 | 0.391 | 0.292 | 0.075 |
| char193 | 0 | 16 | < 17 | -1.469 | -1.484 | -1.485 | -1.443 | -1.362 | -0.542 | -0.189 |
| char193 | 1 | 17 | < 18 | -1.328 | -1.317 | -1.317 | -1.248 | -1.180 | -0.346 | -0.186 |
| char193 | 2 | 18 | < 19 | -1.238 | -1.264 | -1.262 | -1.193 | -1.174 | -0.276 | -0.186 |
| char193 | 3-<18 | 19 | < 20 | -0.096 | -0.086 | -0.086 | -0.091 | -0.074 | -0.067 | 0.001 |
| char193 | 18-High | 20 | | 0.039 | 0.038 | 0.038 | 0.037 | 0.032 | 0.004 | 0.001 |
| char193 | NO INFORMATION | 21 | " = 0 " | 0.000 | 0.000 | 0.000 | 0.000 | 0.000 | 0.000 | 0 |
| char211 | -9999999 | 22 | | -0.144 | -0.096 | -0.096 | -0.103 | -0.092 | -0.219 | -0.002 |
| char211 | -9999998 | 23 | | 0.394 | 0.545 | 0.545 | 0.551 | 0.540 | 0.556 | 0.095 |
| char211 | 0 | 24 | < 31 | 0.514 | 0.449 | 0.449 | 0.453 | 0.434 | 0.195 | 0.061 |
| char211 | 1-<7 | 25 | < 30 | 0.025 | -0.064 | -0.064 | -0.066 | -0.061 | -0.055 | -0.003 |
| char211 | 7-<35 | 26 | < 27 | -0.874 | -0.917 | -0.916 | -0.923 | -0.903 | -0.892 | -0.158 |
| char211 | 35-<80 | 27 | < 28 | -0.629 | -0.674 | -0.674 | -0.677 | -0.658 | -0.634 | -0.106 |
| char211 | 80-<200 | 28 | < 29 | -0.364 | -0.392 | -0.392 | -0.396 | -0.386 | -0.389 | -0.065 |
| char211 | 200-<400 | 29 | < 30 | -0.003 | -0.064 | -0.064 | -0.066 | -0.061 | -0.055 | -0.012 |
| char211 | 400-<800 | 30 | < 31 | 0.026 | -0.064 | -0.064 | -0.066 | -0.061 | -0.055 | -0.003 |
| char211 | 800-<1300 | 31 | < 32 | 0.515 | 0.449 | 0.449 | 0.453 | 0.434 | 0.381 | 0.061 |
| char211 | 1300-<1700 | 32 | < 33 | 0.516 | 0.449 | 0.449 | 0.453 | 0.434 | 0.381 | 0.061 |
| char211 | 1700-High | 33 | | 0.516 | 0.449 | 0.449 | 0.453 | 0.434 | 0.521 | 0.061 |
| char211 | NO INFORMATION | 34 | " = 0 " | 0.000 | 0.000 | 0.000 | 0.000 | 0.000 | 0.000 | 0 |
| char260 | -9999999 | 35 | | -0.138 | -0.162 | -0.162 | -0.164 | -0.160 | -0.173 | -0.028 |
| char260 | 0-<101 | 36 | > 37 | 0.429 | 0.464 | 0.463 | 0.466 | 0.461 | 0.489 | 0.085 |
| char260 | 101-<210 | 37 | > 38 | 0.275 | 0.314 | 0.314 | 0.318 | 0.309 | 0.344 | 0.051 |
| char260 | 210-<305 | 38 | > 39 | 0.275 | 0.312 | 0.312 | 0.312 | 0.305 | 0.258 | 0.051 |
| char260 | 305-<565 | 39 | > 40 | 0.113 | 0.143 | 0.143 | 0.144 | 0.132 | 0.121 | 0.013 |
| char260 | 565-<700 | 40 | > 41 | -0.366 | -0.276 | -0.276 | -0.271 | -0.267 | -0.159 | -0.044 |
| char260 | 700-High | 41 | | -0.367 | -0.276 | -0.276 | -0.271 | -0.267 | -0.159 | -0.044 |
| char260 | NO INFORMATION | 42 | " = 0 " | 0.000 | 0.000 | 0.000 | 0.000 | 0.000 | 0.000 | 0 |

| Char | Attribute | Att. # | Constraint | INFORM-NLP | GEF Max Div | Score S1 | Weights S2 | S3 | S4 | S5 |
|------|-----------|--------|-----------|-----------|------------|------|---------|----|----|----|
| char320 | -9999999-<0 | 43 | > 44 | 0.246 | 0.366 | 0.366 | 0.355 | 0.364 | 0.174 | 0.088 |
| char320 | 0-<590 | 44 | > 45 | 0.171 | 0.105 | 0.105 | 0.108 | 0.107 | 0.121 | 0.021 |
| char320 | 590-<2055 | 45 | > 46 | 0.171 | 0.105 | 0.105 | 0.108 | 0.107 | 0.121 | 0.021 |
| char320 | 2055-<8405 | 46 | > 47 | -0.113 | -0.088 | -0.088 | -0.087 | -0.086 | -0.082 | -0.019 |
| char320 | 8405-<16960 | 47 | > 48 | -0.186 | -0.120 | -0.120 | -0.124 | -0.127 | -0.135 | -0.032 |
| char320 | 16960-<20000 | 48 | > 49 | -0.187 | -0.120 | -0.120 | -0.124 | -0.127 | -0.135 | -0.032 |
| char320 | 20000-<30000 | 49 | > 50 | -0.279 | -0.213 | -0.213 | -0.219 | -0.214 | -0.195 | -0.032 |
| char320 | 30000-<40375 | 50 | > 51 | -0.476 | -0.361 | -0.361 | -0.370 | -0.371 | -0.286 | -0.059 |
| char320 | 40375-<70000 | 51 | > 52 | -0.566 | -0.361 | -0.361 | -0.370 | -0.371 | -0.401 | -0.076 |
| char320 | 70000-High | 52 |  | -0.567 | -0.361 | -0.361 | -0.370 | -0.371 | -0.412 | -0.076 |
| char320 | NO INFORMATION | 53 | " = 0 " | 0.000 | 0.000 | 0.000 | 0.000 | 0.000 | 0.000 | 0 |
| char330 | 0 | 54 | > 55 | 0.236 | 0.251 | 0.251 | 0.253 | 0.249 | 0.293 | 0.044 |
| char330 | 1-<250 | 55 | > 56 | -0.143 | -0.144 | -0.144 | -0.146 | -0.142 | -0.174 | -0.024 |
| char330 | 250-High | 56 |  | -0.311 | -0.343 | -0.343 | -0.344 | -0.341 | -0.392 | -0.062 |
| char330 | NO INFORMATION | 57 | " = 0 " | 0.000 | 0.000 | 0.000 | 0.000 | 0.000 | 0.000 | 0 |
| char380 | -9999999-<0 | 58 |  | 0.000 | 0.000 | 0.000 | 0.000 | 0.000 | 0.000 | 0 |
| char380 | 0-<635 | 59 | > 60 | 0.094 | 0.106 | 0.106 | 0.105 | 0.106 | 0.087 | 0.024 |
| char380 | 635-<1210 | 60 | > 61 | -0.001 | -0.014 | -0.014 | -0.014 | -0.024 | -0.033 | -0.017 |
| char380 | 1210-<1915 | 61 | > 62 | -0.001 | -0.014 | -0.014 | -0.014 | -0.024 | -0.033 | -0.017 |
| char380 | 1915-<5000 | 62 | > 63 | -0.304 | -0.332 | -0.332 | -0.330 | -0.335 | -0.292 | -0.079 |
| char380 | 5000-High | 63 |  | -0.731 | -0.775 | -0.775 | -0.765 | -0.731 | -0.497 | -0.103 |
| char380 | NO INFORMATION | 64 | " = 0 " | 0.000 | 0.000 | 0.000 | 0.000 | 0.000 | 0.000 | 0 |
| char471 | -9999999 | 65 |  | 0.000 | 0.000 | 0.000 | 0.000 | 0.000 | 0.000 | 0 |
| char471 | 0 | 66 | < 67 | -0.417 | -0.429 | -0.429 | -0.426 | -0.402 | -0.250 | -0.039 |
| char471 | 1-<101 | 67 |  | 0.015 | 0.016 | 0.016 | 0.016 | 0.015 | 0.009 | 0.002 |
| char471 | NO INFORMATION | 68 | " = 0 " | 0.000 | 0.000 | 0.000 | 0.000 | 0.000 | 0.000 | 0 |
| char503 | 0 | 69 |  | 0.010 | 0.010 | 0.010 | 0.009 | 0.009 | 0.004 | 0.002 |
| char503 | 1-High | 70 | < 69 | -1.468 | -1.482 | -1.482 | -1.401 | -1.355 | -0.547 | -0.219 |
| char503 | NO INFORMATION | 71 | " = 0 & < 69 " | 0.000 | 0.000 | 0.000 | 0.000 | 0.000 | 0.000 | 0 |
| char533 | -9999999-<1 | 72 |  | 0.132 | 0.156 | 0.156 | 0.154 | 0.150 | 0.107 | 0.025 |
| char533 | 1 | 73 | > 74 | -0.342 | -0.359 | -0.359 | -0.361 | -0.356 | -0.351 | -0.065 |
| char533 | 2 | 74 | > 75 | -0.712 | -0.850 | -0.849 | -0.841 | -0.815 | -0.497 | -0.13 |
| char533 | 3 | 75 | >76 | -0.712 | -0.910 | -0.909 | -0.893 | -0.859 | -0.497 | -0.131 |
| char533 | 4 | 76 | >77 | -0.713 | -0.910 | -0.909 | -0.893 | -0.859 | -0.497 | -0.131 |
| char533 | 5-High | 77 |  | -0.714 | -0.910 | -0.909 | -0.893 | -0.859 | -0.497 | -0.131 |
| char533 | NO INFORMATION | 78 | " = 0 " | 0.000 | 0.000 | 0.000 | 0.000 | 0.000 | 0.000 | 0 |
| char635 | 0 | 79 |  | 0.000 | 0.004 | 0.004 | 0.002 | 0.002 | -0.024 | 0.001 |
| char635 | 1-<3 | 80 | > 81 | 0.071 | 0.050 | 0.050 | 0.055 | 0.053 | 0.105 | 0.007 |
| char635 | 3 | 81 | > 82 | 0.071 | 0.050 | 0.050 | 0.055 | 0.053 | 0.105 | 0.007 |
| char635 | 4 | 82 | > 83 | -0.201 | -0.153 | -0.153 | -0.144 | -0.139 | -0.053 | -0.027 |
| char635 | 5 | 83 | > 84 | -0.474 | -0.400 | -0.400 | -0.376 | -0.389 | -0.122 | -0.091 |
| char635 | 6-High | 84 |  | -0.652 | -0.619 | -0.619 | -0.601 | -0.563 | -0.285 | -0.091 |
| char635 | NO INFORMATION | 85 | " = 0 " | 0.000 | 0.000 | 0.000 | 0.000 | 0.000 | 0.000 | 0 |
| char665 | 0 | 86 | > 87 | 0.105 | 0.106 | 0.106 | 0.106 | 0.105 | 0.094 | 0.02 |
| char665 | 1 | 87 | >88 | -0.532 | -0.529 | -0.529 | -0.531 | -0.523 | -0.483 | -0.096 |
| char665 | 2-High | 88 |  | -0.552 | -0.572 | -0.572 | -0.573 | -0.567 | -0.483 | -0.107 |
| char665 | NO INFORMATION | 89 | " = 0 " | 0.000 | 0.000 | 0.000 | 0.000 | 0.000 | 0.000 | 0 |

| Char | Attribute | Att. # | Constraint | INFORM-NLP | GEF Max Div | S1 | S2 | S3 | S4 | S5 |
|---|---|---|---|---|---|---|---|---|---|---|
| char710 | -9999999 | 90 | " = 0 " | 0.000 | 0.000 | 0.000 | 0.000 | 0.000 | 0.000 | 0 |
| char710 | -9999998 | 91 | " = 0 " | 0.000 | 0.000 | 0.000 | 0.000 | 0.000 | 0.000 | 0 |
| char710 | 0 | 92 | > 93 | 0.072 | 0.066 | 0.066 | 0.066 | 0.062 | 0.056 | 0.007 |
| char710 | 1-<360 | 93 | > 94 | 0.072 | 0.066 | 0.066 | 0.066 | 0.062 | 0.056 | 0.007 |
| char710 | 360-<675 | 94 | >95 | 0.013 | 0.043 | 0.043 | 0.041 | 0.036 | 0.001 | 0.003 |
| char710 | 675-<2435 | 95 | > 96 | -0.359 | -0.343 | -0.343 | -0.342 | -0.320 | -0.272 | -0.035 |
| char710 | 2435-High | 96 | | -0.359 | -0.343 | -0.343 | -0.342 | -0.320 | -0.272 | -0.035 |
| char710 | NO INFORMATION | 97 | " = 0 " | 0.000 | 0.000 | 0.000 | 0.000 | 0.000 | 0.000 | 0 |
| char830 | 0 | 98 | > 99 | 0.010 | 0.010 | 0.010 | 0.009 | 0.009 | 0.004 | 0.002 |
| char830 | 1 | 99 | > 100 | 0.008 | 0.009 | 0.009 | 0.009 | 0.009 | 0.004 | 0.002 |
| char830 | 2-High | 100 | | -0.347 | -0.352 | -0.353 | -0.335 | -0.325 | -0.131 | -0.052 |
| char830 | NO INFORMATION | 101 | " = 0 " | 0.000 | 0.000 | 0.000 | 0.000 | 0.000 | 0.000 | 0 |
| char835 | 0 | 102 | . 103 | 0.054 | 0.056 | 0.056 | 0.054 | 0.053 | 0.035 | 0.01 |
| char835 | 1 | 103 | >104 | -0.336 | -0.339 | -0.339 | -0.337 | -0.338 | -0.250 | -0.062 |
| char835 | 2 | 104 | > 105 | -0.395 | -0.420 | -0.420 | -0.405 | -0.377 | -0.250 | -0.062 |
| char835 | 3 | 105 | > 106 | -0.556 | -0.566 | -0.566 | -0.529 | -0.515 | -0.250 | -0.093 |
| char835 | 4-High | 106 | | -0.875 | -0.910 | -0.911 | -0.829 | -0.811 | -0.250 | -0.15 |
| char835 | NO INFORMATION | 107 | " = 0 " | 0.000 | 0.000 | 0.000 | 0.000 | 0.000 | 0.000 | 0 |
| char840 | 0 | 108 | > 109 | 0.186 | 0.235 | 0.235 | 0.222 | 0.216 | 0.084 | 0.036 |
| char840 | 1 | 109 | > 110 | -0.183 | -0.313 | -0.313 | -0.283 | -0.274 | -0.001 | -0.042 |
| char840 | 2 | 110 | > 111 | -0.937 | -1.062 | -1.062 | -1.027 | -1.006 | -0.595 | -0.175 |
| char840 | 3-High | 111 | | -1.389 | -1.497 | -1.497 | -1.449 | -1.419 | -0.786 | -0.248 |
| char840 | NO INFORMATION | 112 | " = 0 " | 0.000 | 0.000 | 0.000 | 0.000 | 0.000 | 0.000 | 0 |
| char843 | 1 | 113 | | 0.738 | 0.737 | 0.737 | 0.720 | 0.693 | 0.447 | 0.096 |
| char843 | 2 | 114 | | -0.200 | -0.188 | -0.188 | -0.213 | -0.201 | -0.495 | -0.025 |
| char843 | 3 | 115 | | 0.011 | 0.001 | 0.001 | 0.013 | 0.005 | 0.165 | -0.012 |
| char843 | 4 | 116 | | -0.010 | -0.013 | -0.013 | -0.002 | -0.002 | 0.123 | 0.001 |
| char843 | NO INFORMATION | 117 | " = 0 " | 0.000 | 0.000 | 0.000 | 0.000 | 0.000 | 0.000 | 0 |
| char860 | 0 | 118 | > 119 | 0.010 | 0.010 | 0.010 | 0.009 | 0.009 | 0.004 | 0.002 |
| char860 | 1-High | 119 | | -0.523 | -0.528 | -0.528 | -0.499 | -0.483 | -0.195 | -0.078 |
| char860 | NO INFORMATION | 120 | " = 0 " | 0.000 | 0.000 | 0.000 | 0.000 | 0.000 | 0.000 | 0 |
| char870 | 0 | 121 | > 122 | 0.010 | 0.010 | 0.010 | 0.009 | 0.009 | 0.004 | 0.002 |
| char870 | 1 | 122 | > 123 | -0.286 | -0.289 | -0.289 | -0.273 | -0.265 | -0.081 | -0.038 |
| char870 | 2-High | 123 | | -0.287 | -0.289 | -0.289 | -0.273 | -0.265 | -0.114 | -0.044 |
| char870 | NO INFORMATION | 124 | " = 0 " | 0.000 | 0.000 | 0.000 | 0.000 | 0.000 | 0.000 | 0 |
| char950 | -9999998-<7011 | 125 | > 126 | 0.627 | 0.659 | 0.659 | 0.664 | 0.641 | 0.633 | 0.098 |
| char950 | 3300-<4901 | 126 | > 127 | 0.350 | 0.324 | 0.324 | 0.325 | 0.329 | 0.264 | 0.07 |
| char950 | Travel | 127 | > 128 | 0.349 | 0.324 | 0.324 | 0.325 | 0.329 | 0.264 | 0.07 |
| char950 | 5511-High | 128 | > 129 | 0.349 | 0.324 | 0.324 | 0.325 | 0.329 | 0.264 | 0.07 |
| char950 | MOTO | 129 | > 130 | 0.050 | -0.007 | -0.007 | -0.009 | -0.012 | -0.027 | -0.004 |
| char950 | 5697-<7995 | 130 | > 131 | -0.026 | -0.079 | -0.079 | -0.073 | -0.069 | -0.027 | -0.012 |
| char950 | 3723-<5945 | 131 | > 132 | -0.027 | -0.079 | -0.079 | -0.073 | -0.069 | -0.027 | -0.012 |
| char950 | 5611-<8000 | 132 | > 133 | -0.030 | -0.079 | -0.079 | -0.073 | -0.069 | -0.027 | -0.012 |
| char950 | 4814-<4830 | 133 | > 134 | -0.030 | -0.079 | -0.079 | -0.073 | -0.069 | -0.027 | -0.012 |
| char950 | 5013-<8100 | 134 | > 135 | -0.174 | -0.157 | -0.157 | -0.165 | -0.161 | -0.185 | -0.028 |
| char950 | Gas | 135 | > 136 | -0.175 | -0.157 | -0.157 | -0.165 | -0.161 | -0.185 | -0.028 |
| char950 | 5655-<5949 | 136 | > 137 | -0.265 | -0.193 | -0.193 | -0.191 | -0.188 | -0.185 | -0.032 |
| char950 | 5300-<5942 | 137 | > 138 | -0.266 | -0.193 | -0.193 | -0.191 | -0.188 | -0.185 | -0.032 |
| char950 | 5815-<5963 | 138 | > 139 | -0.455 | -0.490 | -0.490 | -0.499 | -0.476 | -0.364 | -0.064 |
| char950 | 5423-<5655 | 139 | | -0.771 | -0.734 | -0.734 | -0.705 | -0.705 | -0.364 | -0.143 |
| char950 | NO INFORMATION | 140 | " = 0 " | 0.000 | 0.000 | 0.000 | 0.000 | 0.000 | 0.000 | 0 |

| Char | Attribute | Att. # | Constraint | ←-------- INFORM-NLP | ---------- GEF Max Div | Score S1 | Weights S2 | --------- S3 | --------- S4 | -------→ S5 |
|------|-----------|--------|------------|---------|---------|--------|--------|--------|--------|--------|
| char960 | Below -1700 | 141 | < 142 | -0.308 | -0.398 | -0.398 | -0.400 | -0.390 | -0.393 | -0.066 |
| char960 | -1700-<-800 | 142 | < 143 | -0.041 | 0.073 | 0.073 | 0.073 | 0.071 | 0.072 | 0.012 |
| char960 | -800-<-450 | 143 | < 144 | -0.041 | 0.073 | 0.073 | 0.073 | 0.071 | 0.072 | 0.012 |
| char960 | " -450-<High " | 144 | | 0.136 | 0.073 | 0.073 | 0.073 | 0.071 | 0.072 | 0.012 |
| char960 | NO INFORMATION | 145 | " = 0 " | 0.000 | 0.000 | 0.000 | 0.000 | 0.000 | 0.000 | 0 |
| char961 | -9999999 | 146 | | -0.122 | -0.211 | -0.211 | -0.215 | -0.195 | -0.261 | -0.001 |
| char961 | -3000-<-1700 | 147 | < 148 | -0.404 | -0.385 | -0.385 | -0.386 | -0.379 | -0.389 | -0.067 |
| char961 | -1700-<-800 | 148 | < 149 | -0.069 | -0.048 | -0.048 | -0.053 | -0.052 | -0.107 | -0.01 |
| char961 | -800-<550 | 149 | < 150 | 0.135 | 0.148 | 0.148 | 0.152 | 0.146 | 0.195 | 0.02 |
| char961 | 550-High | 150 | | 0.205 | 0.186 | 0.186 | 0.188 | 0.179 | 0.202 | 0.022 |
| char961 | NO INFORMATION | 151 | " = 0 " | 0.000 | 0.000 | 0.000 | 0.000 | 0.000 | 0.000 | 0 |
| char962 | Below -1500 | 152 | < 153 | -0.126 | -0.085 | -0.085 | -0.089 | -0.095 | -0.176 | -0.024 |
| char962 | -1500-<-1100 | 153 | < 154 | -0.060 | -0.085 | -0.085 | -0.089 | -0.095 | -0.174 | -0.024 |
| char962 | -1100-<-850 | 154 | < 155 | -0.054 | -0.073 | -0.073 | -0.074 | -0.072 | -0.094 | -0.012 |
| char962 | -850-<-550 | 155 | < 156 | -0.054 | -0.073 | -0.073 | -0.074 | -0.072 | -0.094 | -0.012 |
| char962 | -550-<-400 | 156 | < 157 | 0.034 | 0.019 | 0.019 | 0.021 | 0.022 | 0.050 | 0.005 |
| char962 | -400-<-300 | 157 | < 158 | 0.035 | 0.019 | 0.019 | 0.021 | 0.022 | 0.050 | 0.005 |
| char962 | -300-<1 | 158 | < 159 | 0.035 | 0.019 | 0.019 | 0.021 | 0.022 | 0.065 | 0.005 |
| char962 | 1-<200 | 159 | < 160 | 0.075 | 0.104 | 0.104 | 0.108 | 0.107 | 0.163 | 0.02 |
| char962 | 200-High | 160 | | 0.123 | 0.174 | 0.174 | 0.177 | 0.180 | 0.223 | 0.041 |
| char962 | NO INFORMATION | 161 | " = 0 " | 0.000 | 0.000 | 0.000 | 0.000 | 0.000 | 0.000 | 0 |
| char965 | Below -950 | 162 | < 163 | -0.329 | -0.328 | -0.328 | -0.329 | -0.327 | -0.330 | -0.065 |
| char965 | -950-<-750 | 163 | < 164 | -0.328 | -0.328 | -0.328 | -0.329 | -0.327 | -0.330 | -0.065 |
| char965 | -750-<-550 | 164 | < 165 | -0.269 | -0.321 | -0.321 | -0.320 | -0.313 | -0.278 | -0.053 |
| char965 | -550-<-400 | 165 | < 166 | -0.269 | -0.321 | -0.321 | -0.320 | -0.313 | -0.278 | -0.053 |
| char965 | -400-<-300 | 166 | < 167 | -0.049 | -0.044 | -0.044 | -0.044 | -0.040 | -0.040 | -0.001 |
| char965 | -300-<-200 | 167 | < 168 | 0.162 | 0.186 | 0.186 | 0.186 | 0.187 | 0.153 | 0.03 |
| char965 | -200-<-100 | 168 | < 169 | 0.192 | 0.201 | 0.201 | 0.203 | 0.187 | 0.191 | 0.03 |
| char965 | -100-<80 | 169 | < 170 | 0.349 | 0.366 | 0.366 | 0.367 | 0.364 | 0.357 | 0.07 |
| char965 | 80-High | 170 | | 0.350 | 0.366 | 0.366 | 0.367 | 0.364 | 0.357 | 0.07 |
| char965 | NO INFORMATION | 171 | " = 0 " | 0.000 | 0.000 | 0.000 | 0.000 | 0.000 | 0.000 | 0 |
| **Development Divergence** | | | | 1.732 | 1.753 | 1.753 | 1.752 | 1.731 | 1.641 | 0.926 |
| **Validation Divergence** | | | | 1.617 | 1.636 | 1.6360 | 1.6361 | 1.628 | 1.594 | 0.856 |